\documentclass[preprint2]{aastex}

\shorttitle{Correlation between broad-line and radio variations}

\shortauthors{Liu, Bai, Wang \& Li}

\begin{document}

\title{Hints of correlation between broad-line and radio variations for 3C 120}

\author{H. T. Liu\altaffilmark{1,2}, J. M. Bai\altaffilmark{1,2}, J. M. Wang\altaffilmark{3,4} and S. K. Li\altaffilmark{1,2}}

\altaffiltext{1}{Yunnan Observatories, Chinese Academy of
Sciences, Kunming, Yunnan 650011, China}

\altaffiltext{2}{Key Laboratory for the Structure and Evolution of
Celestial Objects, Chinese Academy of Sciences, Kunming, Yunnan
650011, China}

\altaffiltext{3}{Key Laboratory for Particle Astrophysics,
Institute of High Energy Physics, Chinese Academy of Sciences, 19B
Yuquan Road, Beijing 100049, China}

\altaffiltext{4}{Theoretical Physics Center for Science
Facilities, Chinese Academy of Sciences, Beijing 100049, China}

\email{htliu@ynao.ac.cn}

\begin{abstract}

In the paper, we investigate correlation between broad-line and
radio variations for broad-line radio galaxy 3C 120. By the
z-transformed discrete correlation function method and the
model-independent flux randomization/random subset selection
(FR/RSS) Monte Carlo method, we find that the broad H$\beta$ line
variations lead the 15 GHz variations. The FR/RSS method shows
that the H$\beta$ line variations lead the radio variations by a
factor of $\tau_{\rm{ob}}=0.34\pm0.01$ yr. This time lag can be
used to locate the position of emitting region of radio outbursts
in jet, on the order of $\sim$ 5 light-years, from the central
engine. This distance is much larger than the size of broad-line
region. The large separation of the radio outburst emitting region
from the broad-line region will observably influence the gamma-ray
emission in 3C 120.

\end{abstract}

\keywords{galaxies: active -- galaxies: individual (3C 120) --
galaxies: jets -- quasars: emission lines -- radio continuum:
galaxies}

\section{INTRODUCTION}

According to the reverberation mapping model \citep[e.g.][]{b9},
the broad emission line variations follow the ionizing continuum
variations through the photoionization process. The variation
correlations between broad-lines and continua were observed with
time lags in type 1 active galactic nuclei \citep[AGNs, see
e.g.][]{b26,b27,b39}. The disturbances from the central engine in
AGNs are transported with ionizing continua to broad-lines.
Theoretical researches show that the jets can be ejected from
inner accretion disk in the vicinity of the central black hole
\citep[e.g.][]{b34,b11,b10,b33}. \citet{b39d} indicated a disk-jet
symbiosis with comparable power channelled through the disk and
the jet. Correlations between radio powers and broad-line
luminosities are found for AGNs and are regarded as evidence for
the disk-jet symbiosis \citep[see
e.g.][]{b13a,b11b,b11c,b50c,b55b}. These previous results indicate
that the disturbances in the central engine are likely propagated
outwards along the jets. Observations show that dips in the X-ray
emission, generated in the central engine, are followed by
ejections of bright superluminal radio knots in the jets of AGNs
and microquasars \citep[e.g.][]{b31,b6}. The dips in the X-ray
emission are well correlated with the ejections of bright
superluminal knots in the radio jets of  3C 120 \citep{b12} and 3C
111 \citep{b13}. The outbursts are physically linked to the
ejections of superluminal knots \citep[e.g.][]{b47}. Then these
outbursts of broad-line and jet emission might respond to the
stronger disturbances in the central engine. It is expected that
there might be correlations with time lags between variations of
broad-line and jet emission. A method was proposed to connect the
time lags, the size of broad-line region (BLR), and the location
of jet emission for blazar 3C 273 \citep[][hereafter Paper
I]{b29}.

The BLRs are important to gamma-ray emission in blazars. The gamma
rays from blazars are generally believed to be from a relativistic
jet with a small viewing angle \citep{b10a}. The diffuse radiation
field of BLR could have a strong impact on the expected external
Compton (EC) spectrum of the most powerful blazars \citep[see
e.g.][]{b40a,b50b,b55a,b39a,b56a,b41a,b41b,b10b,b41c}. This strong
impact rises from two factors. One of them is the seed photons
from the BLR in the inverse Compton scattering, and the seed
photons significantly influence the EC spectrum. The other is
photon-photon absorption between the seed photons and the
gamma-ray photons of the EC spectrum. There is a underlying
physical factor that constrains how much the above two factors
influence on the gamma-ray spectrum. The underlying factor is the
location of gamma-ray--emitting region relative to the BLR. If the
gamma-ray--emitting region is inside the BLR, the gamma-ray
spectrum will shift to higher energies and the gamma-ray
luminosity will become larger due to the relativistic effects. At
the same time, the photon-photon absorption becomes more
significant as the emitting region goes deeper into the BLR
\citep[see][]{b55a,b56a,b10b}. As the emitting region is outside
the BLR, the gamma-ray spectrum will shift to lower energies and
the gamma-ray luminosity will become lower due to the relativistic
effects. At the same time, the photon-photon absorption becomes
insignificant as the emitting region keeps away from the BLR. Thus
it is valuable to connect the BLR size with the location of jet
emission. It will be an important step for this connection to
confirm correlation between variations of broad-lines and radio
emission of jet and to estimate the relevant time lags. In the
paper, we study this issue in the broad-line radio galaxy 3C 120.

The structure of this paper is as follows. Section 2 presents
method. Section 3 presents application to 3C 120. Section 3
contains three subsections: subsection 3.1 presents constraint on
time lag, subsection 3.2 data of 3C 120, and subsection 3.3
analysis of time lag. Section 4 is for discussion and conclusions.
In this work, we assume the standard $\Lambda$CDM cosmology with
$H_0=70 \rm{\/\ km \/\ s^{-1} \/\ Mpc^{-1}}$, $\Omega_{\rm{M}}$ =
0.27, and $\Omega_{\rm{\Lambda}}$= 0.73.

\section{METHOD}
According to equation (7) in Paper I, we have a relation between
$R_{\rm{BLR}}$, $R_{\rm{radio}}$, and $\tau_{\rm{ob}}$
\begin{equation}
R_{\rm{BLR}}=R_{\rm{radio}} \left(\frac{c}{v_{\rm{d}}}-\cos \theta
\right )-\frac{c \langle \tau_{\rm{ob}}\rangle}{1+z},
\end{equation}
where $R_{\rm{BLR}}$ is the size of the BLR, $R_{\rm{radio}}$ is
the radio-emitting location of the jet, $c$ is the speed of light,
$v_{\rm{d}}$ is the travelling speed of disturbances down the jet,
equivalent to the bulk velocity of jet $v_{\rm{j}}$, $\theta$ is
the viewing angle of the jet axis to the line of sight, and
$\langle \tau_{\rm{ob}}\rangle \equiv \tau_{\rm{ob}}$ is the
measured time lag of the radio emission relative to the broad
lines. From the velocity $\beta=v_{\rm{j}}/c$ and the viewing
angle $\theta$, we have the apparent speed $\beta_{\rm{a}}=\beta
\rm{sin\theta} /(1-\beta \rm{cos\theta})$, which gives $\beta =
\beta_{\rm{a}}/(\beta_{\rm{a}} \rm{cos\theta}+\rm{sin\theta})$.
Substituting this expression of $\beta$ for the velocity term in
equation (1), we have
\begin{equation}
R_{\rm{BLR}}=R_{\rm{radio}}\frac{\sin \theta}{\beta_{\rm{a}}}
-\frac{c \langle \tau_{\rm{ob}}\rangle}{1+z}.
\end{equation}
From equation (2), we have an expression to estimate $\langle
\tau_{\rm{ob}}\rangle$
\begin{equation}
\langle \tau_{\rm{ob}}\rangle=\left(R_{\rm{radio}}\frac{\sin
\theta}{\beta_{\rm{a}}} - R_{\rm{BLR}}\right)\frac{1+z}{c}.
\end{equation}

\section{APPLICATION TO 3C 120}
3C 120, at redshfit $z=0.033$, has the one-sided jet with the
apparent superluminal motion in the approaching jet \citep{b20b}.
The central supermassive black hole of this object has a mass of a
few times $10^7 M_{\odot}$ \citep{b35,b38}.

\subsection{Constraint on time lag}
The Very Long Baseline Array (VLBA) imaging observations revealed
a very complex radio jet in 3C 120, such as superluminal
components, stationary components, and trailing components in
inner jet \citep{b20b,b22,b12,b28b}. There are two stationary
features in the radio jet, D and S1 \citep{b28b}. The feature D is
located at the base of the jet, and could be the core of the jet.
The feature S1 is most likely a standing shock formed in the jet.
As the moving knots pass through the stationary feature S1, these
knots will produce outbursts. The process was tested in the
optical light curves of 3C 120 with peaks corresponding to the
moving component passages through S1. It is not possible to verify
whether the radio flux density reacted to the passage of moving
features in the same fashion as the optical continuum. The
prominent superluminal feature o reaches it maximum flux density
around 0.5 mas from the radio core \citep{b20b}. The stationary
feature S1 was identified at $\sim$ 0.7 mas separated from the
VLBA 43 GHz core \citep{b28b}. Thus the radio outbursts may
originate in the region from D to S1 in the inner jet.

For 3C 120, \citet{b12} derived a distance of $\sim$ 0.5 pc from
the corona to the VLBA 43 GHz core region using the average time
delay between the start of the X-ray dips and the time of ejection
of the corresponding superluminal knots. For 3C 111, a distance of
$\sim$ 0.6 pc from the corona is estimated with the same method as
in 3C 120 \citep{b13}. The 43 GHz VLBA observations give the
global parameters of the jet for 3C 120, such as $\theta = 20.5\pm
1.8^{\circ}$\citep{b22}. These global parameters are widely
accepted by other researches. The 15 GHz components may correspond
to the strongest events in the central engine \citep{b28b}. Due to
the optical depth, the 15 GHz outburst may follow the 43 GHz
outburst. The emitting location of the 15 GHz outburst is
estimated as the sum of the distance of 0.5 pc from the corona to
the VLBA core and the de-projected distance of 0.0--0.7 mas from
the VLBA core. Under the standard $\Lambda$CDM cosmology we
considered, an angular separation of 1 mas in the sky corresponds
to a projected linear distance of 0.66 pc. The de-projected
distance of 0.7 mas from the radio core is equal to
0.7$\times$0.66 pc/$\rm{sin\theta}$=0.7$\times$0.66 pc/$\rm{sin
20.5^{\circ}}$=1.32 pc. The distance of 15 GHz outburst emitting
region from the central engine is $R_{\rm{radio}}\sim $0.5 pc--0.5
pc+1.32 pc=0.5--1.82 pc=1.63--5.93 ly. For 3C 120, the H$\beta$
line has a BLR size of $R_{\rm{BLR}}=43.8^{+27.7}_{-20.3}$
light-days, i.e. $R_{\rm{BLR}}=0.12^{+0.08}_{-0.06}$ ly
\citep{b35}. The apparent speeds of the moving components with
well-determined motions are all within a range of $\beta_{\rm{a}}=
4.0 \pm 0.2$ \citep{b12}. Based on equation (3),
$R_{\rm{BLR}}=0.12$ ly, $\beta_{\rm{a}}= 4.0$, $\theta =
20.5^{\circ}$ and $R_{\rm{radio}}$=1.63--5.93 ly, we derive
$\tau_{\rm{ob}}>0$. This positive time lag means that the
broad-line variations lead the radio variations.

\subsection{Data of 3C 120}
We make use of the 15 GHz light curve with a higher sampling rate
of 59 times per year. This radio light curve is published in
\citet{b39b}. For the H$\beta$ line, \citet{b33a} presents a light
curve with a very dense sampling of 20 times per month, and also
\citet{b20a} presents a light curve of sampling 20 times per month
in the reverberation mapping observations. These light curves are
presented in Figure 1, and are used to analyze the
cross-correlation between them.

\begin{figure}[htp]
\begin{center}
\includegraphics[angle=-90,scale=0.25]{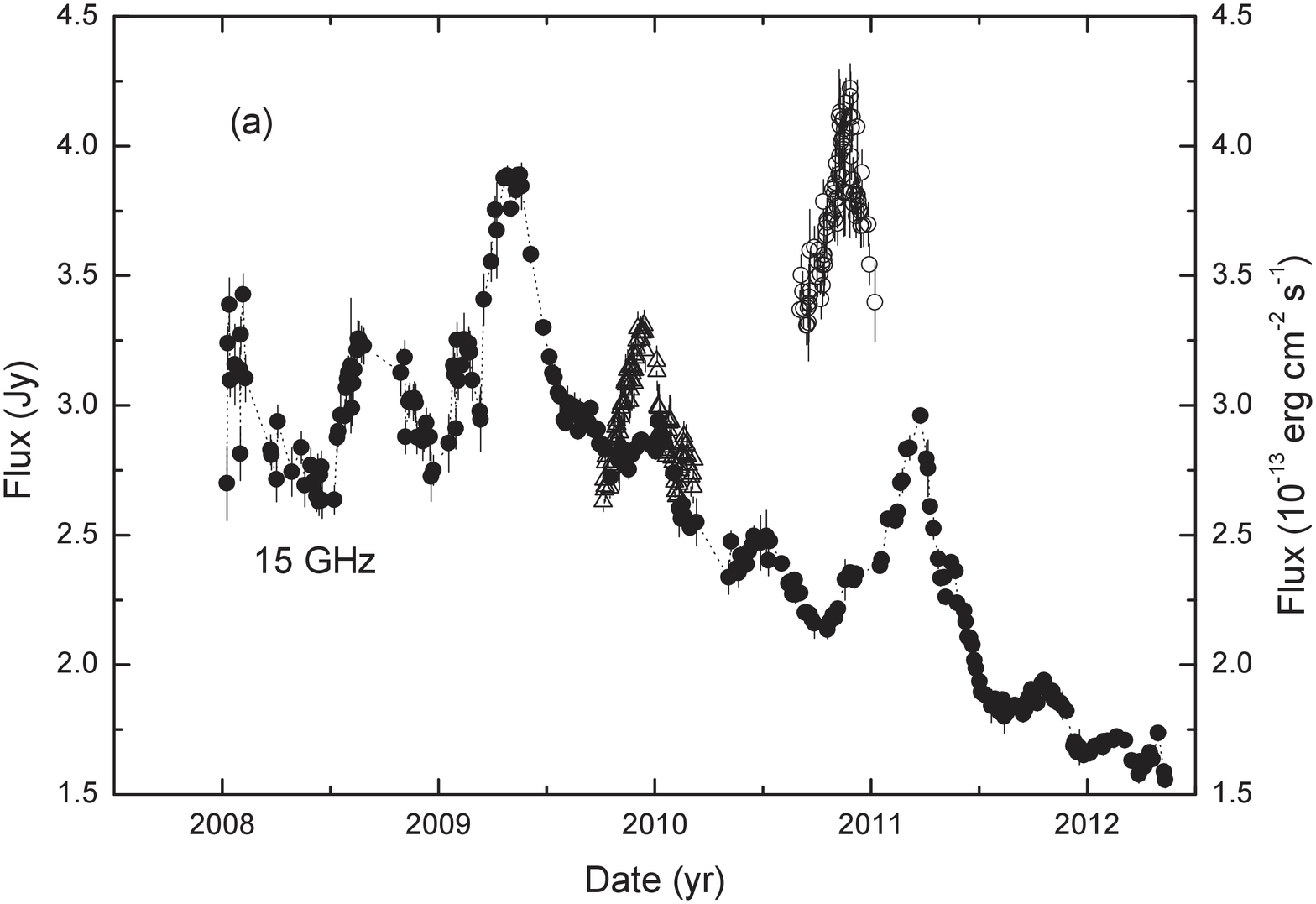}
\includegraphics[angle=-90,scale=0.25]{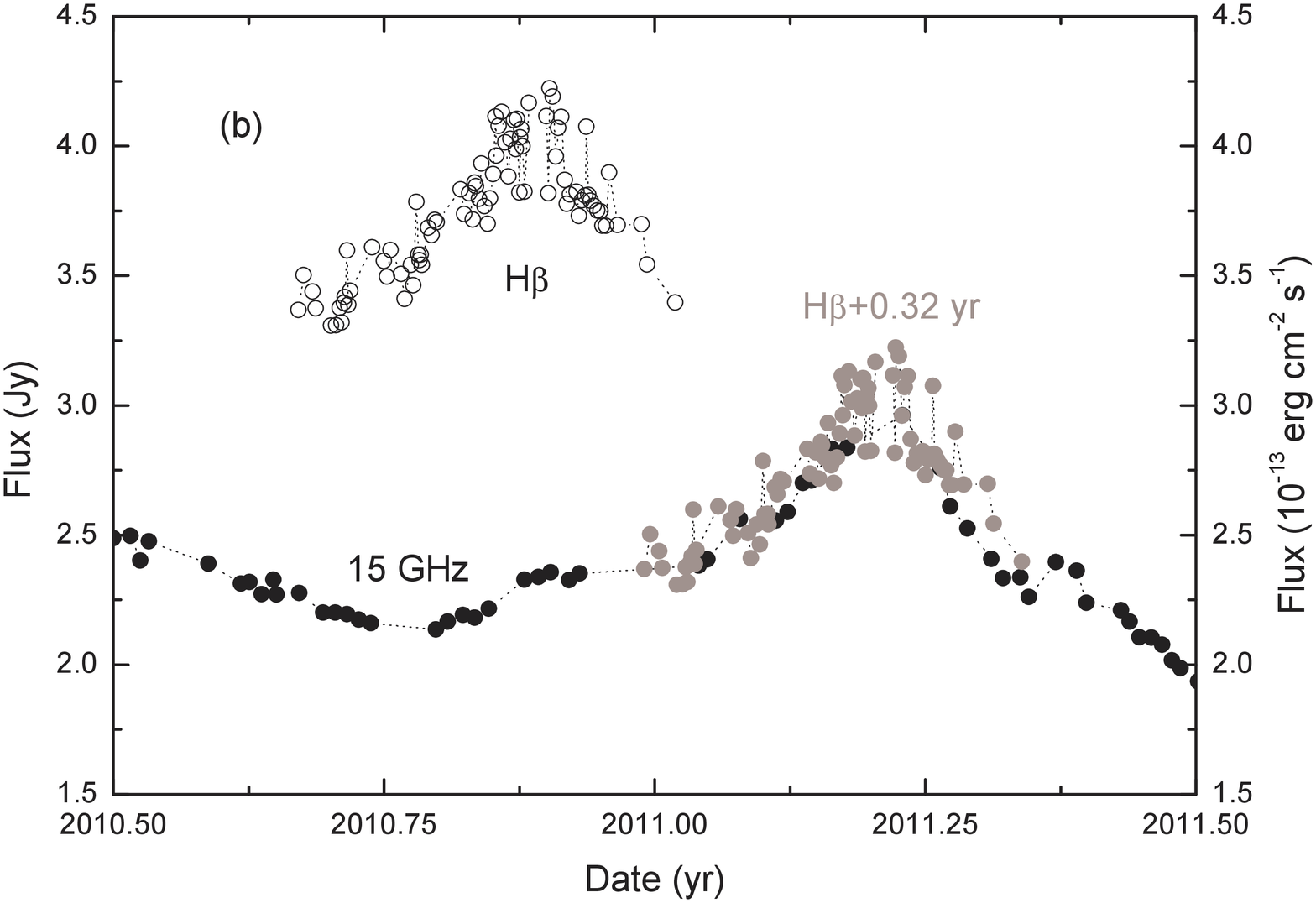}

\end{center}
 \caption{Light curves of H$\beta$ and 15 GHz emission. Black open triangles denote the H$\beta$ light curve of \citet{b33a}.
   Black open circles denote the H$\beta$ light curve of \citet{b20a}. Black solid circles denote the 15 GHz light curves in units of Jy.
  The H$\beta$ line is in units of $10^{-13}\rm{\/\ erg \/\ cm^{-2}\/\ s^{-1}}$. Gray solid circles denote the H$\beta$ light curve moved along the $x$- and $y$-axes.}
  \label{fig1}
\end{figure}

\subsection{Analysis of Time Lag}
The z-transformed discrete correlation function
\citep[ZDCF;][]{b5} is used to analyze time lags characterized by
the centroid of the ZDCF. The ZDCF method is straightforward to
determine whether there is a time lag between different light
curves, and firstly it is applied to analyze the time lags. The
centroid time lag $\tau_{\rm{cent}}$ is computed by all the points
with correlation coefficients not less than 0.8 times the maximum
of correlation coefficients in the ZDCF bumps closer to the
zero-lag. The uncertainties of each point in the ZDCFs only take
into account the uncertainties from the measurements by Monte
Carlo simulation, and the uncertainties of time lags are
underestimated \citep[see][hereafter Paper II]{b30}. Thus we use
the model-independent flux randomization/random subset selection
(FR/RSS) Monte Carlo method \citep{b36} to re-estimate the time
lags and their uncertainties in the cross-correlation results. The
FR/RSS method is based on the discrete correlation function (DCF)
method \citep{b15} for the sparsely sampled light curves, and on
the interpolated cross-correlation function (ICCF) method for the
densely sampled light curves.

The H$\beta$ line light curves with a very dense sampling are
published in 2012 \citep{b20a,b33a}. 3C 120 is densely observed
from 2008.0 to 2012.5 in the 15 GHz radio monitoring program with
the 40 m telescope at the Owens Valley Radio Observatory
\citep{b39b}. Firstly, it is obvious that the H$\beta$ line light
curve in \citet{b20a} can be well matched with the outburst in the
15 GHz light curve from 2011.0 to 2011.3 as the line light curve
is moved right by 0.32 yr (see Figure 1b). The ZDCF method and the
FR/RSS method are performed to investigate the correlation between
these H$\beta$ line light curves and the 15 GHz light curve, and
estimate the time lags from the correlation. As these two line
light curves are combined into one light curve, the calculated
ZDCF is presented in Figure 2a. The horizontal and vertical error
bars in Figure 2a represent the 68.3 per cent confidence intervals
in the time lags and the relevant correlation coefficients,
respectively. There are positive and negative correlations (see
Figure 2a). The positive correlation has a time lag around 0.3 yr.
The H$\beta$ line variations lead the 15 GHz variations by a
factor of $\sim$ 0.3 yr. This lag has the same sign as that lag
estimated in section 3.1. This indicates that the positive
correlation is reliable. For the positive correlation, the ZDCF
method gives $\tau_{\rm{cent}}=0.305^{+0.011}_{-0.002}$ yr. The
FR/RSS method gives $\tau_{\rm{cent}}=0.336^{+0.012}_{-0.010}$ yr
with a mean of peak correlation coefficients $r=0.59\pm0.07$ in
Monte Carlo simulations of 10,000 runs (see Figure 2b). This time
lag is well consistent with that lag derived from the ZDCF method.
\begin{figure}[htp]
 \begin{center}
 \includegraphics[angle=-90,scale=0.25]{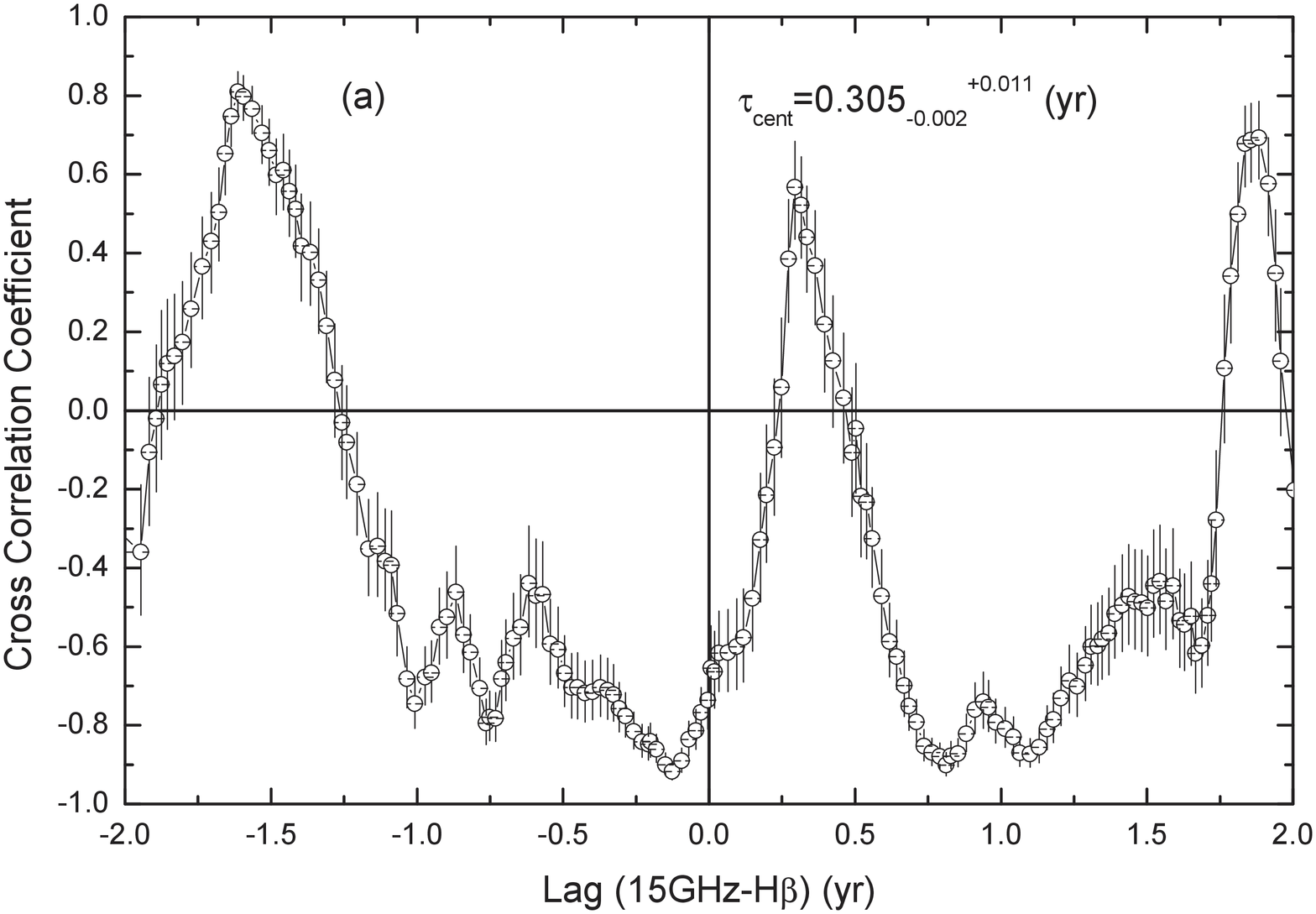}
 \includegraphics[angle=-90,scale=0.25]{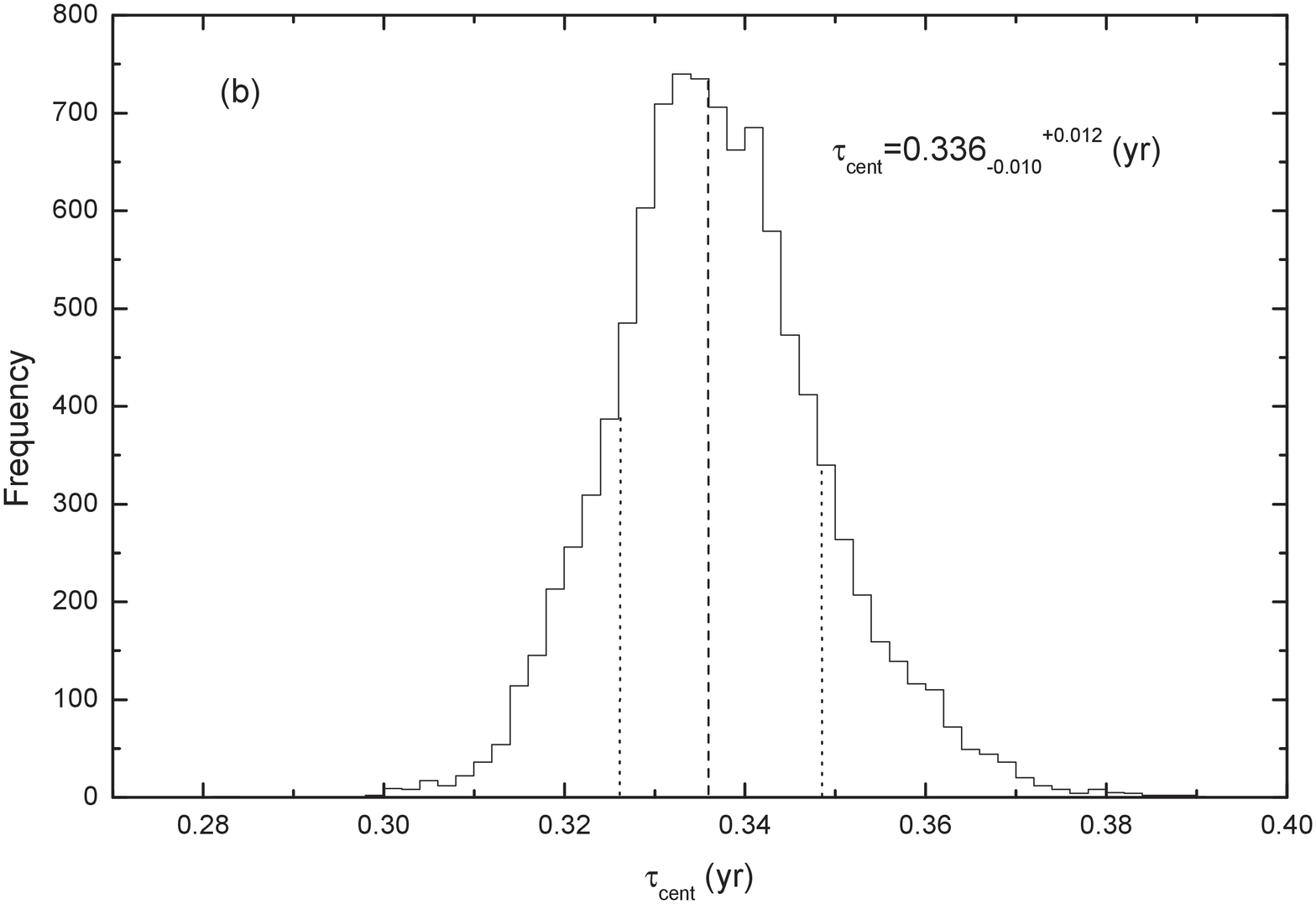}
 \end{center}
 \caption{(a) ZDCF between the H$\beta$ and 15 GHz light curves.
 (b) Distribution of $\tau_{\rm{cent}}$ obtained with the FR/RSS method.
 The vertical dashed line is the median of distribution, and the dotted
 lines show the 68.3 per cent confidence interval of $\tau_{\rm{cent}}$.}
  \label{fig2}
\end{figure}

The 15 GHz light curve shows a simple baseline superimposed with
some outbursts and flares (see Figure 3a). This baseline should
influence the cross-correlation function between this radio light
curve and the H$\beta$ line light curve. Thus we subtract this
baseline assumed as a simple gaussian profile from the 15 GHz
light curve. It is obvious that the simple gaussian profile can
well account for the underlying baseline in the 15 GHz light curve
(see Figure 3a). The cross-correlation function between the
residual radio light curve and the H$\beta$ line light curve is
calculated with the ZDCF method and the FR/RSS method. The
calculated ZDCF is presented in Figure 3b. The positive
cross-correlation around 0.3 yr is re-confirmed, and the
significance of cross-correlation is significantly improved with
this residual light curve. The ZDCF method gives
$\tau_{\rm{cent}}=0.328^{+0.014}_{-0.004}$ yr. The FR/RSS method
gives $\tau_{\rm{cent}}=0.34\pm0.01$ yr with a mean of peak
correlation coefficients $r=0.88\pm0.02$ in Monte Carlo
simulations of 10,000 runs (see Figure 3c). This time lag is in
excellent agreement with that lag derived from the ZDCF method.
The broad H$\beta$ line variations lead the 15 GHz variations.
Hereafter, $\tau_{\rm{cent}}$ is equivalent to $\tau_{\rm{ob}}$.
\begin{figure}[htp]
\begin{center}
\includegraphics[angle=-90,scale=0.25]{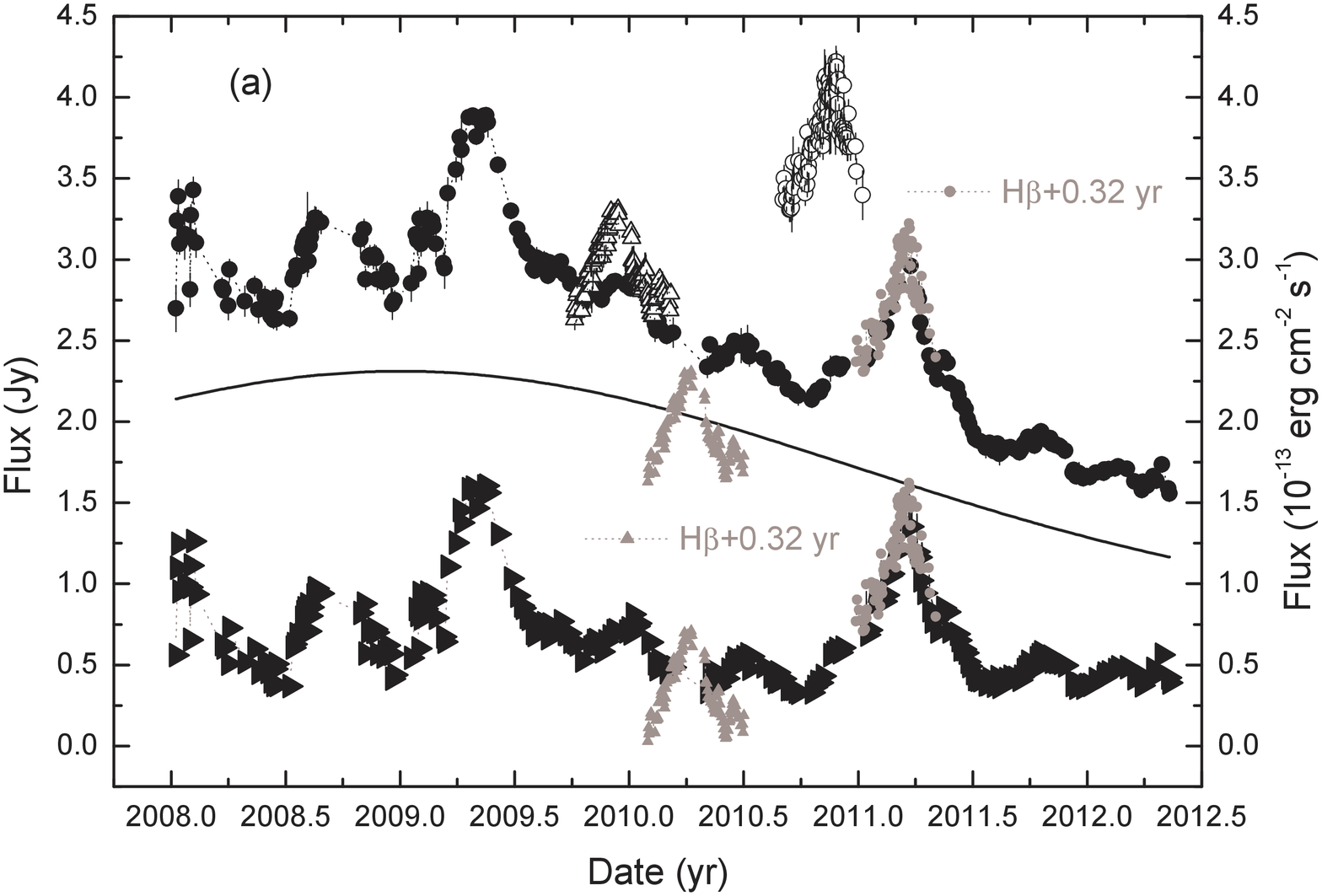}
\includegraphics[angle=-90,scale=0.25]{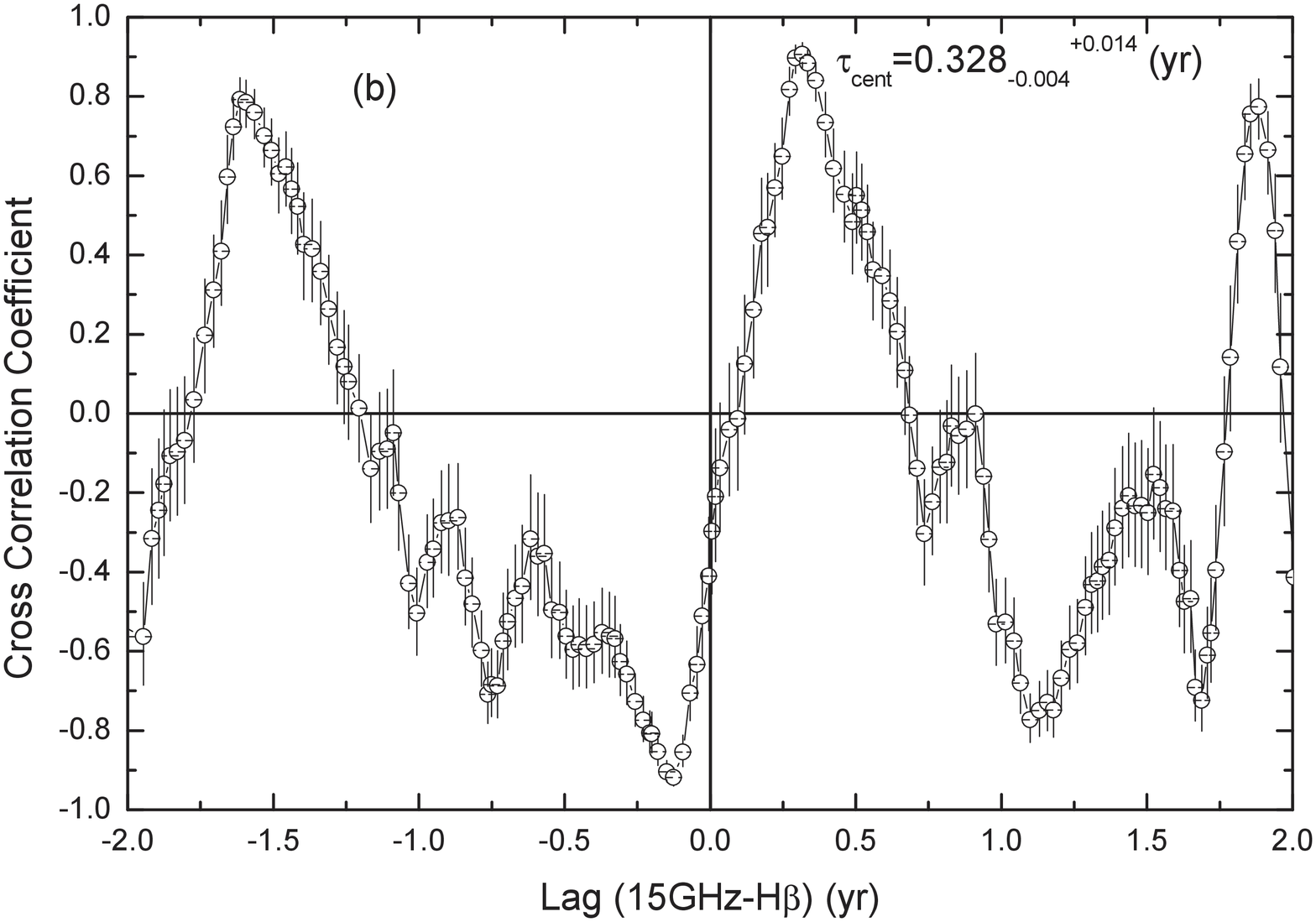}
\includegraphics[angle=-90,scale=0.25]{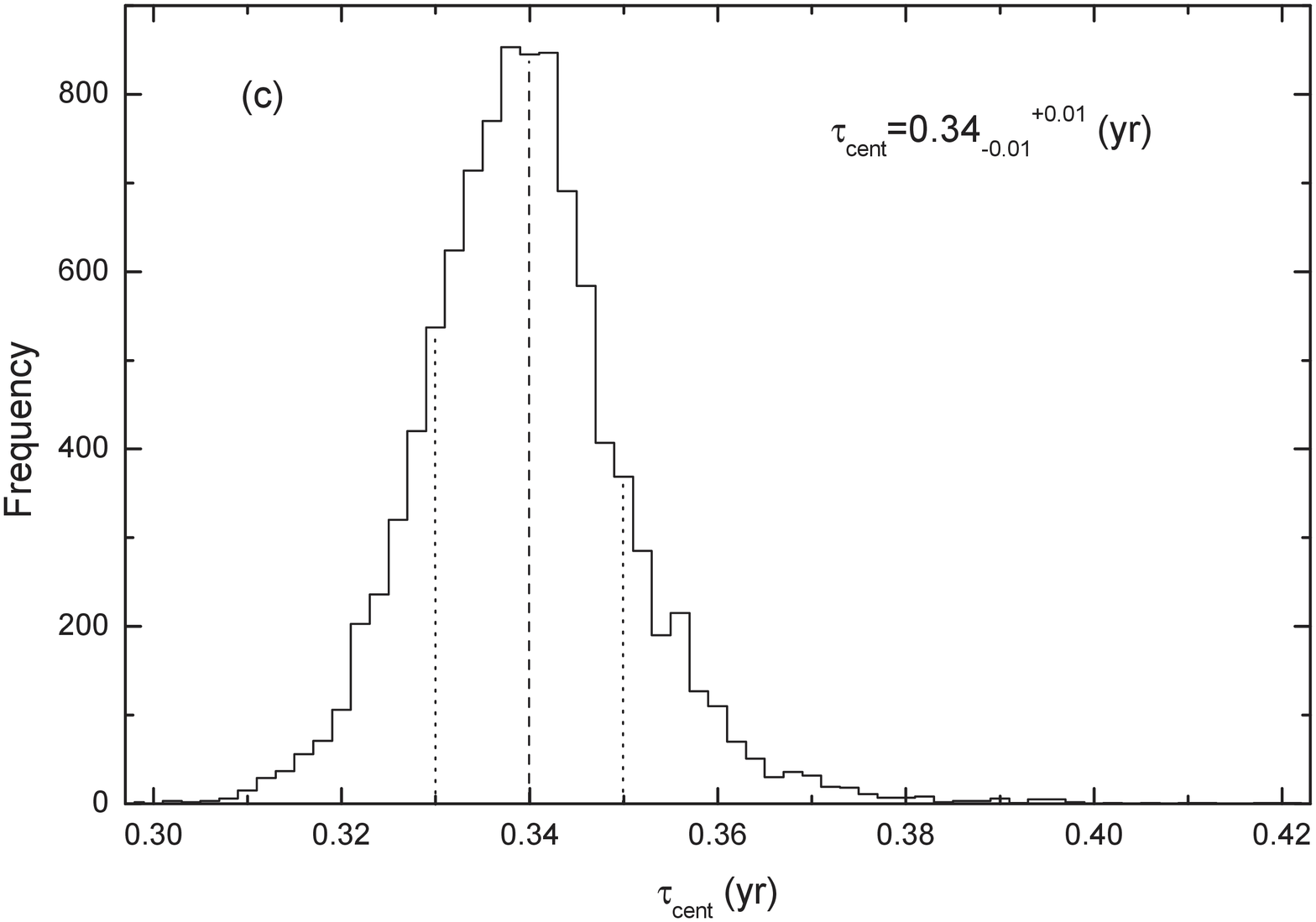}
\end{center}
 \caption{(a) Comparison between the 15 GHz and H$\beta$ light curves. Black right triangles are the 15 GHz light curve subtracted by
 a assumed simple baseline denoted with the black solid line. Gray symbols are the H$\beta$ light curve moved along the $x$- and $y$-axes.
 Other symbols are same as in Figure 1. (b) ZDCF between the modified 15 GHz light curve and the H$\beta$ light curve in black color.
 (c) Distribution of $\tau_{\rm{cent}}$ obtained with the FR/RSS method. The vertical dashed and dotted
 lines are same as in Figure 2b.}
  \label{fig3}
\end{figure}

\section{DISCUSSION AND CONCLUSIONS}

We simplify the ionizing continuum region to be a point. This
simplification indicates that the disturbances will simultaneously
be transported outwards with the ionizing continuum and the
relativistic jet, i.e. it makes equations (1)--(3) to be valid.
This simplification will influence the time lag $\tau_{\rm{ob}}$.
The disturbances in the accretion disk will take a certain time to
travel between their location of origin and the event horizon of
the central black hole. Then the disturbances will take some time
to pass through the ionizing continuum region to the event
horizon. For 3C 120, the UV ionizing continuum region is located
at $\sim$ 5$r_{\rm{g}}$ from the black hole, where
$r_{\rm{g}}=GM_{\rm{BH}}/c^2$ is the gravitational radius of the
black hole \citep{b12}. The black hole mass is on the order of
$10^7 M_{\odot}$ \citep{b35,b38}, and the size of $\sim$
5$r_{\rm{g}}$ is on the order of $10^{-5}$ ly. If the disturbances
are a thermal fluctuation propagating inward, it should have an
effective speed $\la 0.01c$ \citep{b12}, to cause a time delay of
$\ga 0.001$ yr for the distance of $\sim$ 5$r_{\rm{g}}$. This time
delay should be negligible compared with the time lag
$\tau_{\rm{ob}}=0.34$ yr. In estimation of the magnitude of
$\tau_{\rm{ob}}$ in section 3.1, we use the distance of $\sim$ 0.5
pc from the VLBA 43 GHz core region to the corona in the accretion
disk-corona system, and ignore the size of corona. The size of
corona may influence $\tau_{\rm{ob}}$. The coronal radius is
$\sim$ 40$r_{\rm{g}}$ \citep{b12}. The jet velocity near the
central engine will be $\sim$ 0.9c (see Paper I), and this radius
of $\sim$ 40$r_{\rm{g}}$ will cause a time delay on the order of
$10^{-4}$ yr. The time delay is negligible.

From equation (2), we have an expression to estimate
$R_{\rm{radio}}$ from $\beta_{\rm{a}}$, $\theta$, $R_{\rm{BLR}}$
and $\tau_{\rm{ob}}$
\begin{equation}
R_{\rm{radio}}=\frac{\beta_{\rm{a}}}{\sin \theta} \left(
R_{\rm{BLR}}+\frac{c \langle \tau_{\rm{ob}}\rangle}{1+z} \right).
\end{equation}
For $\beta_{\rm{a}}=4.0\pm 0.2$, $\theta = 20.5\pm 1.8^{\circ}$,
$R_{\rm{BLR}}=0.12^{+0.08}_{-0.06}$ ly, and $\tau_{\rm{ob}}=0.34
\pm 0.01$ yr, we have $R_{\rm{radio}}=5.24 \pm 0.16$ ly from Monte
Carlo simulations based on equation (4). Thus we have
$R_{\rm{BLR}}\ll R_{\rm{radio}}$ for 3C 120. $\it{Fermi}$-Large
Area Telescope (LAT) detected gamma rays from 3C 120 \citep{b27a},
and it was suggested that the GeV emission of broad-line radio
galaxies is most likely dominated by the beamed radiation of
relativistic jets observed at intermediate viewing angles. The
radio and gamma-ray emitting regions are closely connected with
each other, and there is $R_{\rm{\gamma}}\la R_{\rm{radio}}$
between the radio-emitting position $R_{\rm{radio}}$ and the
gamma-ray--emitting location $R_{\rm{\gamma}}$
\citep[e.g.,][]{b14,b21,b28,b41,b2}. It is unclear for 3C 120 that
$R_{\rm{\gamma}}\la R_{\rm{BLR}}$ or
$R_{\rm{\gamma}}>R_{\rm{BLR}}$, which will significantly influence
the gamma-ray spectrum produced in the EC processes. The locations
of gamma-ray--emitting regions relative to the BLRs are still an
open and controversial issue in the researches on blazars. There
are three options for the issue. The first option is that
$R_{\rm{\gamma}}\la R_{\rm{BLR}}$ for the powerful blazars
\citep[e.g.][]{b16,b56a,b10b,b41c,b18}. The second one is that
$R_{\rm{\gamma}}> R_{\rm{BLR}}$
\citep[e.g.][]{b7,b28a,b41d,b40,b32,b52,b51}. The third option is
that the same source can display both behaviors. That is, most of
the time the dissipation region is inside the BLR, but there could
be some epochs when the gamma-ray--emitting region drift outside
the BLR. The very first idea was advanced in \citet{b15a}, and a
very clear case with multi-wavelength coverage was recently found
by \citet{b18a}. The gamma-ray light curves and the corresponding
broad-line light curves should shed light on this issue.

The chosen parameters are average for the jet, and do not
correspond to any of the components identified in \citet{b22} and
\citet{b12}. The jet components will have different orientations
and different velocities. This seems to be an issue for the choice
of jet parameters in equations (1)--(4). The reverberation mapping
model assumes the linear response of broad emission lines to
ionizing continuum. In fact, the line response is not linear. It
is most likely that the line and radio emission respond
nonlinearly to the events in the central engine. Thus some weaker
events in the central engine might not produce the correlative
responses in the radio and broad-line variations. Both radio and
broad-line emission may have good responses to the strongest
events in the central engine, and their relevant outbursts should
have good matching. There is a matching between these outbursts in
the H$\beta$ line and 15 GHz light curves sampled densely.
However, the overall complexity of the light curves and the jet
structure at radio bands may lead to difficulties of
cross-identifying individual events in different bands. Thus it is
difficult to identify the radio knots and radio outbursts
corresponding to the broad-line outbursts. We investigated the
correlation and time lag between the radio and broad-line light
curves by comparing their profiles and cross-correlating them.

The 15 GHz light curve is observed with the 40 m telescope at the
Owens Valley Radio Observatory. The 40 m telescope can not resolve
the inner jet on the pc scales. Then the 15 GHz fluxes contain all
the emission from the inner jet. It is difficult to identify the
component responsible for the 15 GHz outburst. It is not possible
to determine the relevant velocities of this component along the
jet from the central engine to the emitting site of outburst. Thus
the average velocity of primary components rather than trailing
features will be a good proxy of the global velocity of component
emitting outburst. The viewing angle is the same case as the
velocity. The similar choice is accepted for the jet parameters
for 3C 273 (see Paper I). There are positive and negative time
lags between radio variations and those of broad-lines H$\alpha$,
H$\beta$, and H$\gamma$ due to the relative short coverage of
these line light curves in 3C 273. The longer ultraviolet line
light curves show that these broad-line variations lag the radio
variations (see Paper II). Thus the broad-line variations lag the
radio variations. A constraint of $R_{\rm{\gamma}}\lesssim$
0.40--2.62 pc is set by these negative time lags (see Paper I).
The gamma-ray flares detected with $\it{Fermi}$-LAT set a limit of
$R_{\rm{\gamma}}<$ 1.6 pc for 3C 273 \citep{b39c}. The limit is
marginally consistent with that constraint of
$R_{\rm{\gamma}}\lesssim$ 0.40--2.62 pc. The acceptance of the
average parameters of inner jet is reasonable.

In this paper, we find correlation between the broad-line and
radio variations with the ZDCF method and the FR/RSS method, and
determine a positive lag for the broad-line radio galaxy 3C 120.
The positive lag means that the 15 GHz variations lag the H$\beta$
line variations. We derive $\tau_{\rm{ob}}=0.34\pm 0.01$ yr from
the FR/RSS method. This time lag is consistent with that estimated
from the ZDCF method. Monte Carlo simulations give the
radio-emitting location $R_{\rm{radio}}=5.24\pm 0.16$ ly from this
time lag, the average parameters of inner jet, and equation (4).
It is reasonable to use the average parameters of inner jet in
equations (1)--(4). This reasonability is supported by the
marginal agreement of our previous constraint of
$R_{\rm{\gamma}}\lesssim$ 0.40--2.62 pc with the limit of
$R_{\rm{\gamma}}<$ 1.6 pc from $Fermi$-LAT observations of
gamma-ray flares in 3C 273. The underlying baseline of 15 GHz
light curve significantly influences the correlation between broad
H$\beta$ line and 15 GHz variations. The subtraction of this
baseline from the 15 GHz light curve can well improve the
correlation. The longer H$\beta$ line light curve well sampled may
test the correlation. The existence of this correlation is a key
to connect the BLR size with the emitting location of jet, and it
is important to the gamma-ray emission of AGNs.

\acknowledgements We are grateful to Dr. L. Foschini for
constructive comments and suggestions. HTL thanks the National
Natural Science Foundation of China (NSFC; Grant 11273052) for
financial support. JMB acknowledges the support of the NSFC (Grant
11133006). HTL thanks the financial support of the Youth
Innovation Promotion Association, CAS and the project of the
Training Programme for the Talents of West Light Foundation, CAS.

\clearpage

\end{document}